\documentclass[a4paper]{jpconf_mod}
\usepackage{graphicx}
\usepackage{amsmath,xfrac,paralist}

\def\t#1{\tilde{ #1}}
\def\tev{\textrm{TeV}}
\def\gev{\textrm{GeV}}
\def\inb{\ensuremath{\textrm{nb}^{-1}}}
\def\ipb{\ensuremath{\textrm{pb}^{-1}}}
\def\ifb{\ensuremath{\textrm{fb}^{-1}}}

\def\W{\ensuremath{W}}
\def\Z{\ensuremath{Z}}
\def\to{\ensuremath{\rightarrow}}
\def\ttb{\ensuremath{t\bar{t}}}
\def\X{\ensuremath{\tilde\chi_1^0}}
\def\R{\emph{R}}

\def\mz{\ensuremath{m_0}}
\def\mh{\ensuremath{m_{\sfrac{1}{2}}}}

\def\met{\ensuremath{E_{\rm T}^{\rm miss}}}

\def\at{\ensuremath{\alpha_{\rm T}}}

\begin{document}
\title{Dark matter searches at LHC}

\author{Vasiliki A Mitsou$^{1,2}$}

\address{$^1$ Instituto de F\'isica Corpuscular (IFIC), CSIC -- Universitat de Val\`encia, \\
Parc Cient\'ific, Apartado de Correos 22085, E-46071, Valencia, Spain}

\address{$^2$ CERN, PH Department, CH-1211 Geneva 12, Switzerland}

\ead{vasiliki.mitsou@ific.uv.es}

\begin{abstract}
Besides Standard Model measurements and other Beyond Standard Model studies, the ATLAS and CMS experiments at the LHC will search for Supersymmetry, one of the most attractive explanation for dark matter. The SUSY discovery potential with early data is presented here together with some first results obtained with 2010 collision data at 7~\tev. Emphasis is placed on measurements and parameter determination that can be performed to disentangle the possible SUSY models and SUSY look-alike and the interpretation of a possible positive supersymmetric signal as an explanation of dark matter.
\end{abstract}

\section{Introduction}\label{sc:intro}

Unveiling the nature of dark matter (DM)~\cite{dm-review} is a quest in both Astrophysics and Particle Physics. Among the list of well-motivated candidates, the most popular particles are cold and weakly interacting, and typically predict missing energy signals at particle colliders. Supersymmetry (SUSY) and models with extra dimensions are theoretical scenarios that inherently provide such a dark matter candidate. The Large Hadron Collider (LHC)~\cite{lhc} currently in operation at CERN in Geneva, Switzerland, is an ideal machine for discovering DM in colliders and exploring both phenomenological as much as purely theoretical aspects aspects of DM.  

The exploration of dark matter is being complemented by other types of searches of particle dark matter: direct detection in low background underground experiments~\cite{direct} and indirect detection of neutrinos, gamma-rays and antimatter with terrestrial and space-borne detectors~\cite{indirect}. Experiments in upcoming colliders, such as the ILC~\cite{ilc} and CLIC~\cite{clic}, are expected to further constraint such models and make a key step in understanding dark matter.  

The structure of this paper is as follows. Section~\ref{sc:dm} provides a brief introduction to dark matter properties as defined by the current cosmological data. In section~\ref{sc:lhc} the features of collider experiments that play a central role in exploring DM are highlighted. In section~\ref{sc:susy} we discuss the strategy for discovering supersymmetry at the LHC, some recent results and prospects for the near future. Studies on methods to constrain dark matter parameters at the LHC, such as particle masses and spins, are reviewed in section~\ref{sc:dmlhc}. Some alternative theoretical models yielding a modified DM density and its implications for SUSY searches are discussed in section~\ref{sc:alt}. The paper concludes with an outlook in section~\ref{sc:out}.

\section{Dark matter evidence}\label{sc:dm}

The nature of the dark sector of our Universe constitutes one of the major mysteries of fundamental physics. According to observations over the past two decades, most of our Universe energy budget consists of unknown entities: $\sim\!23\%$ is dark matter and $\sim\!72\%$ is dark energy, a form of ground-state energy. Dark energy is believed to be responsible for the current-era acceleration of the Universe. Dark matter, on the other hand, is matter inferred to exist from gravitational effects on visible matter, but is undetectable by emitted or scattered electromagnetic radiation. A possible explanation then ---other than the introduction of a new particle--- is to ascribe the observed effects to modified Newtonian dynamics (MOND)~\cite{mond}. There is a variety of theoretical proposals predicting DM particles interacting weakly only, as discussed in detail below, however the possibility of charged dark matter still remains open~\cite{khlopov}.

\begin{figure}[ht]
\begin{minipage}[b]{0.4\textwidth}
\includegraphics[width=\textwidth]{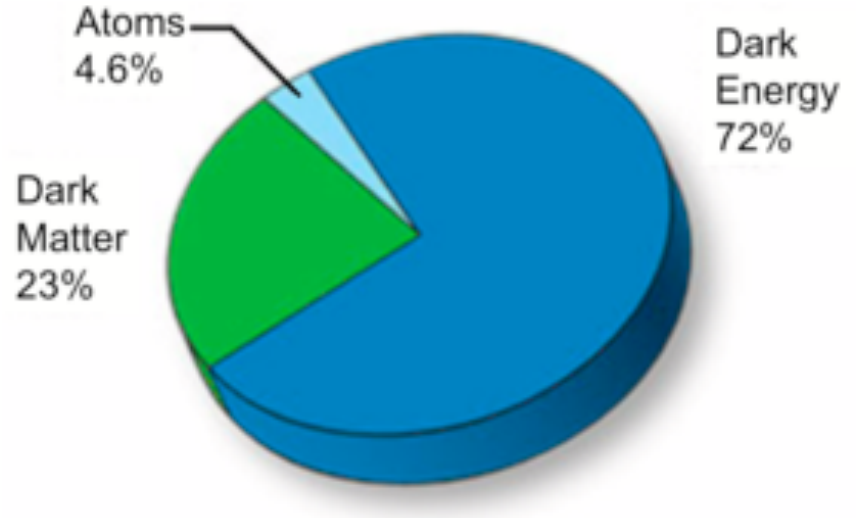}
\caption{\label{fg:budget}The energy distribution of the Universe according to recent cosmological evidence and assuming the $\Lambda$CDM model~\cite{lcdm}.}
\end{minipage}\hspace{2pc}%
\begin{minipage}[b]{0.55\textwidth}
\includegraphics[width=\textwidth]{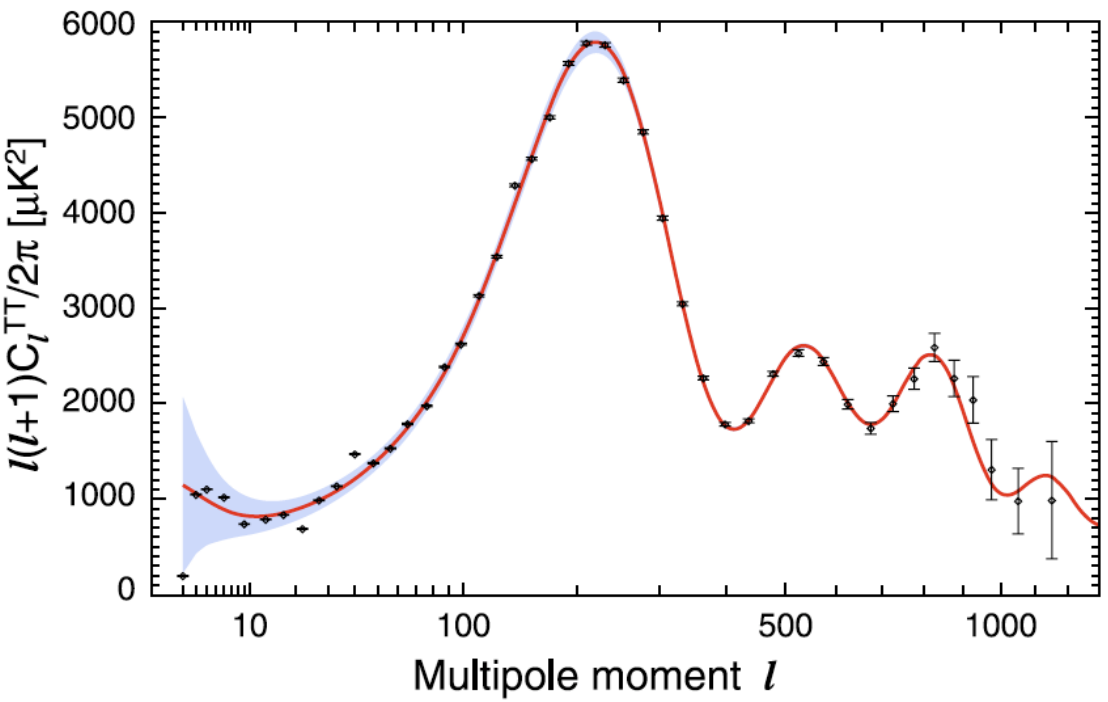}
\caption{\label{fg:cmb}Temperature power spectrum from WMAP7. The third acoustic peak is sensitive to the dark matter density~\cite{cmb}.}
\end{minipage} 
\end{figure}

The energy budget of the Cosmos (fig.~\ref{fg:budget}) has been obtained by combining a variety of astrophysical data, such as type-Ia supernovae~\cite{snIa}, cosmic microwave background (CMB)~\cite{wmap}, baryon oscillations~\cite{bao} and weak lensing data~\cite{lensing}. The most precise measurement comes from anisotropies of the cosmic microwave background~\cite{wmap,cmb}. The third peak in the temperature power spectrum, shown in fig.~\ref{fg:cmb}, can be used to extract information about the dark matter contribution to the Universe energy budget. 

Evidence from the formation of large-scale structure (galaxies and their clusters) strongly favour cosmologies where non-baryonic DM is entirely composed of cold dark matter (CDM), i.e.\ non-relativistic particles.\footnote{The interpretation of the `raw' astrophysical data in the context of the Universe energy/matter content as presented above is based on the Standard Cosmological Model ($\Lambda$CDM)~\cite{lcdm}, involving cold DM as the dominant DM species, and a positive cosmological constant $\Lambda>0$. Nevertheless, as discussed in section~\ref{sc:alt}, the possibility for different theoretical scenarios is open, that modify the estimated DM relic abundance for given cosmological observations.} CDM particles, in turn, may be axions~\cite{axion}, superheavy non-thermal relics (wimpzillas, cryptons)~\cite{shdm} or weakly interacting massive particles (WIMPs). The latter class of DM candidates arises naturally in models which attempt to explain the origin of electroweak symmetry breaking and this is precisely where the connection between Cosmology and Particle Physics lies. Furthermore the typical (weak-scale) cross sections characterizing these models are of the same order of magnitude as the WIMP annihilation cross section, thus establishing the so-called \emph{WIMP miracle}. 

\section{WIMPs and colliders}\label{sc:lhc}

WIMP dark matter candidates include the lightest neutralino in models with weak-scale supersymmetry~\cite{susy}, while Kaluza-Klein photons arise in scenarios with universal extra dimensions (UED)~\cite{ued}, and lightest $T$-odd particles are predicted in Little Higgs models~\cite{little} with a conserved $T$-parity. The common denominator in these theories is that they all predict the existence of a electrically neutral, colourless and \emph{stable} particle, whose decay is prevented by a kind of symmetry: \R-parity, connected to baryon and lepton number conservation in SUSY models; $KK$-parity, the four-dimensional remnant of momentum conservation in the extra dimensions; and a $Z_2$ discrete symmetry called $T$-parity in Little Higgs models. The origin of DM can be attributed to more than one particles, even within the same theoretical framework, e.g.\ in the degenerate scenario of the next-to-minimal supersymmetric standard model (NMSSM)~\cite{grigoris}.

In this paper, we focus on SUSY signatures, although these may be very similar to the other afore-mentioned models. \R-parity is defined as: $R = (-1)^{3(B-L)+2S}$, where $B$, $L$ and $S$ are the baryon number, lepton number and spin, respectively. Hence $R=+1$ for all Standard Model particles and $R=-1$ for all SUSY particles. It is stressed that the conservation of \R-parity is an \emph{ad hoc} assumption. The only firm restriction comes from the proton lifetime: non-conservation of both $B$ and $L$ leads to rapid proton decay. \R-parity conservation has serious consequences in SUSY phenomenology in colliders: the SUSY particles are produced in pairs and the lightest SUSY particle (LSP) is absolutely stable. 

A parenthesis is due here addressing the issue of (not necessarily cold) dark matter in SUSY models with \R-parity violation (RPV)~\cite{rpv}. These seemingly incompatible concepts \emph{can} be reconciled in models with a gravitino~\cite{rpv-grav} or an axino~\cite{rpv-axino} LSP with a lifetime exceeding the age of the Universe. In both cases, RPV is induced by bilinear terms in the superpotential that can also explain current data on neutrino masses and mixings without invoking any GUT-scale physics. Decays of the next-to-lightest superparticle occur rapidly via RPV interaction, and thus they do not upset the Big-Bang nucleosynthesis, unlike the \R-parity conserving case. \R-violating couplings can be sufficiently large to lead to interesting expectations for collider phenomenology; these will be the standard signatures of \R-parity conserving supersymmetry, with multi-lepton or multi-jet events and the possibility of explicit lepton number violation at the final state~\cite{lola}. Nevertheless determining whether \R-parity is conserved or broken may not be trivial as WIMPs, whether absolutely stable or quasi-stable, cannot be detected directly in collider experiments.  

Indeed weakly interacting massive particles do not interact neither electromagnetically nor hadronically with matter and thus, once produced, they traverse the various detectors layers without leaving a trace (just like neutrinos do). However by exploiting the hermeticity of the experiments, we can get a hint of the WIMP presence through the balance of the energy/momentum measured in the various detector components, the so-called \emph{missing energy}. In hadron colliders, in particular, since the (longitudinal) momenta of the colliding partons are unknown, only the  \emph{transverse missing energy}, \met, can be reliably used to `detect' DM particles. 

\section{SUSY(-like) searches}\label{sc:susy}

Here we focus on searches for dark matter in the two general-purpose experiments of the LHC, namely ATLAS~\cite{atlas-det} and CMS~\cite{cms-det}. We review recent studies\footnote{As of December 2010.} performed on up to 300~\inb\ of LHC proton-proton collisions data at a centre-of-mass energy of 7~\tev. For comprehensive accounts of the envisaged analyses, based on Monte Carlo simulations, the reader is referred to the ATLAS `CSC Book'~\cite{atlas-csc} and the CMS Physics Technical Design Report~\cite{cms-tdr}.  

\begin{figure}[hbt]
\begin{minipage}[b]{0.43\textwidth}
\includegraphics[width=\textwidth]{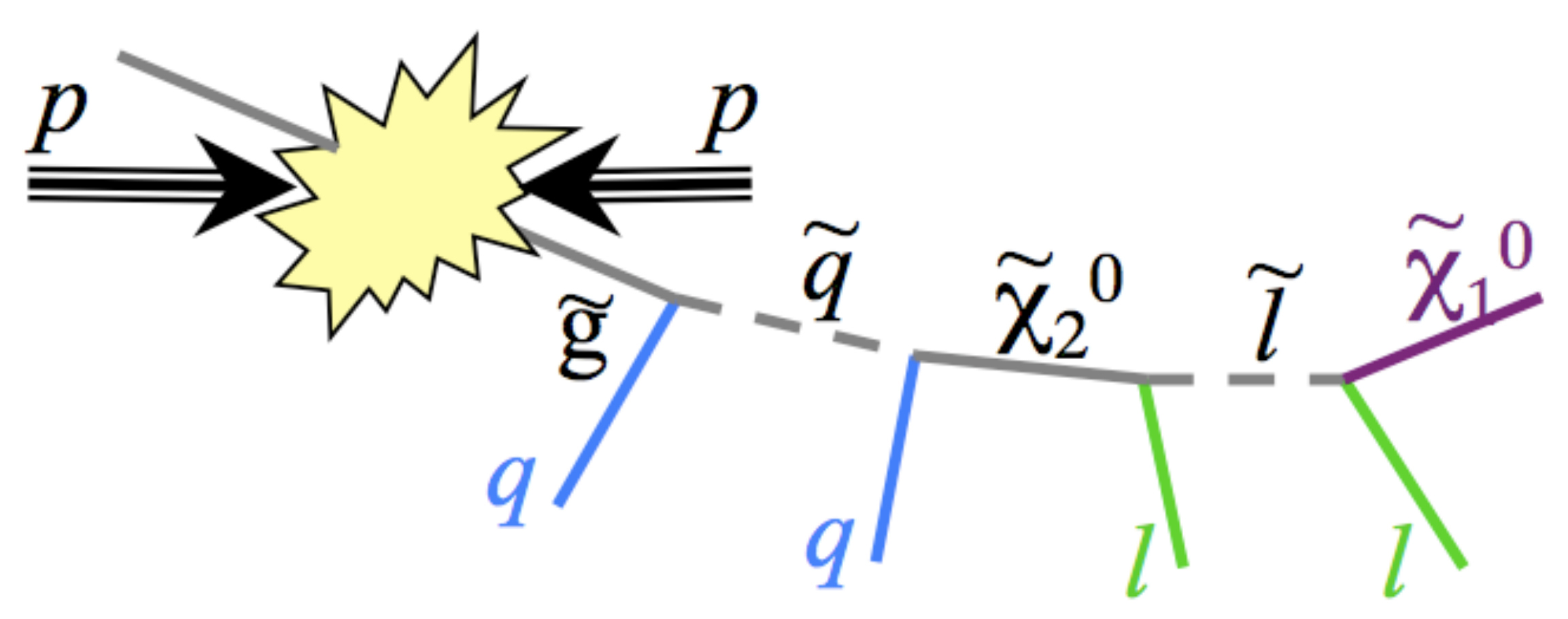}
\caption{\label{fg:cascade}An example of the cascade decay of heavier to lightest sparticles. For simplicity only one branch is shown here; a second one is expected to be present. Note the high lepton and jet multiplicity.}
\begin{eqnarray*}
\met &\equiv& \left\Vert -\sum_{i}\vec{p}_{T,i} \right\Vert \\
&=& \left| -\sum_{i}p_{x,i} - \sum_{i}p_{y,i} \right|
\end{eqnarray*}
\caption{\label{fg:met}\met\ definition: the sums are over all reconstructed objects and calorimeter clusters/towers.}
\end{minipage}\hspace{2pc}%
\begin{minipage}[b]{0.51\textwidth}
\includegraphics[width=\textwidth]{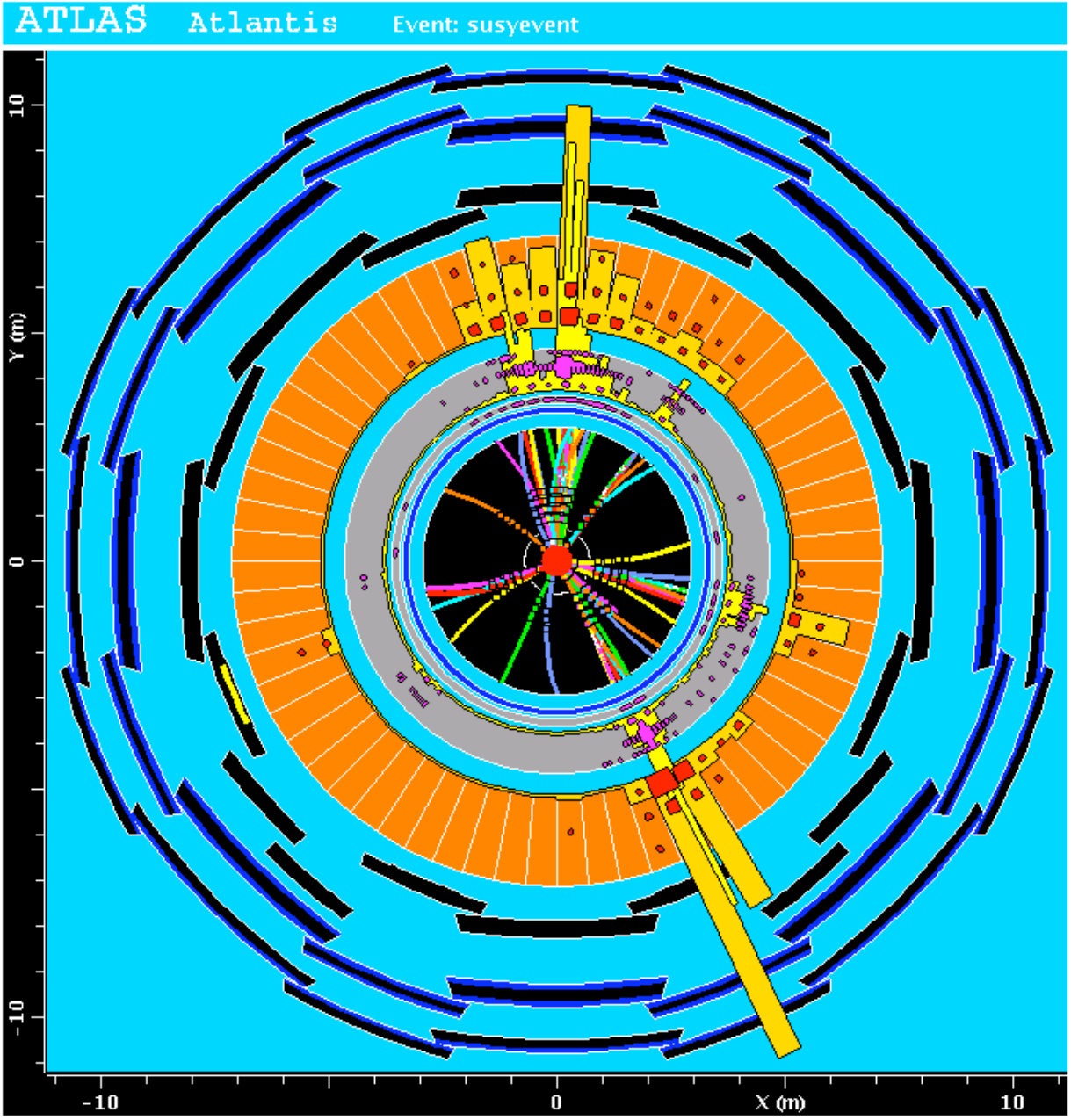}
\caption{\label{fg:etmiss}Transverse view of a simulated SUSY event in the ATLAS detector~\cite{alan}. Note the imbalance of the deposited energy distribution towards the right-hand side. Two (stable) lightest neutralinos (not shown) escape detection towards the left-hand side.}
\end{minipage} 
\end{figure}

At the LHC, supersymmetric particles are expected to be predominantly produced hadronically, i.e.\ through gluino-pair, squark-pair and squark-gluino production. Each of these (heavy) sparticles is going to decay into lighter ones in a cascade decay that finally leads to an LSP, which in most of the scenarios considered is the lightest neutralino \X, as depicted in fig.~\ref{fg:cascade}. The two LSPs would escape detection giving rise to high transverse missing energy, which is rigorously defined in fig.~\ref{fg:met}. Such a simulated event as it will appear in the ATLAS detector is illustrated in fig.~\ref{fg:etmiss}.

The search strategy followed in the inclusive channels is based on the detection of high \met, many jets and possibly energetic leptons. The analyses make extensive use of data-driven Standard Model background measurements. Detailed studies have been carried out for various signatures using Monte Carlo data-sets fully simulated with Geant4~\cite{geant4} for specific SUSY signal parameters and for the relevant Standard Model backgrounds. Although various SUSY-breaking mechanisms have been considered by the two collaborations, the early results and projections for higher integrated luminosity highlighted here have been performed in the context of the minimal Supergravity (mSUGRA) model. By assumption the mSUGRA model avoids both flavour-changing neutral currents and extra sources of $CP$ violation. For masses in the TeV range, it typically predicts too much cold dark matter unless something enhances the \X\ annihilation, as discussed in section~\ref{sc:alt}. The specific points mentioned here, denoted SU$n$ for ATLAS and LM$n$ for CMS, have been chosen to be roughly consistent with the observed CDM and represent a variety of different decay modes. 

\begin{figure}[ht]
\begin{minipage}[b]{0.47\textwidth}
\includegraphics[width=\textwidth]{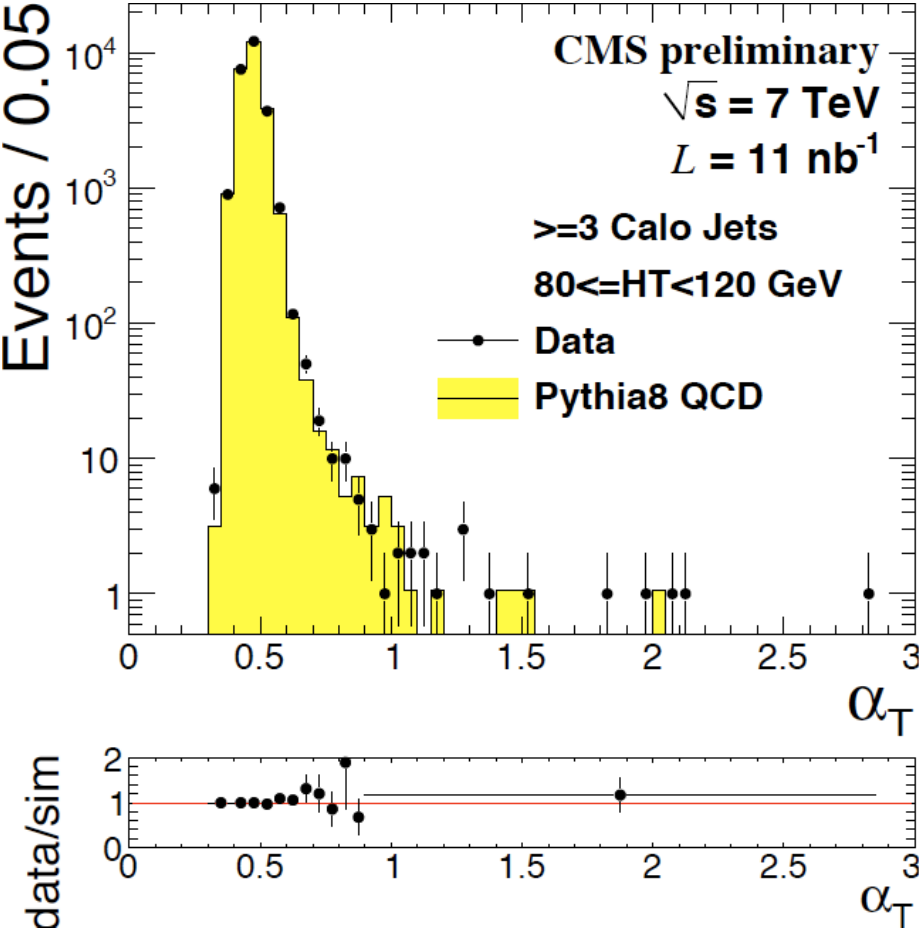}
\caption{\label{fg:cms-at}$a_{\rm T}$ distribution in CMS multijet events for $80 < H_{\rm T} < 120~\gev$. The $a_{\rm T}$ tail is reduced to practically zero (not shown) for $H_{\rm T}>120~\gev$~\cite{cms-at}.}
\end{minipage}\hspace{2pc}%
\begin{minipage}[b]{0.47\textwidth}
\includegraphics[width=\textwidth]{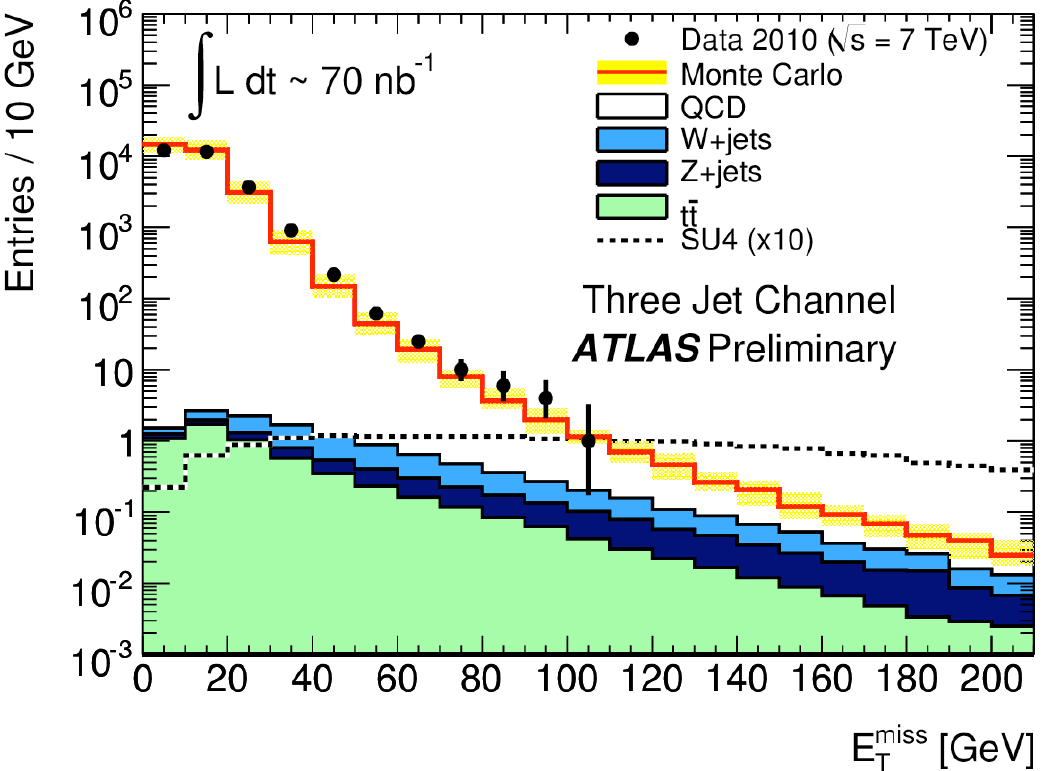}
\caption{\label{fg:atlas-met}Distribution of the missing transverse momentum for ATLAS events in the 0-lepton, three-jet channel after basic preselection~\cite{atlas-met}. Note the long tail in the prediction for the SU4 mSUGRA point.}
\end{minipage} 
\end{figure}

SUSY searches require careful control over backgrounds from standard model processes. Several methods for data-driven background determinations were developed and tested on early LHC $pp$ collision data. Such a method allowed CMS to study QCD backgrounds, to control jet-energy mismeasurement, and to measure background contributions from processes producing non-prompt leptons or hadrons misidentified as leptons~\cite{cms-at}. It is based on the \at\ discriminating variable for the all-hadronic channel, defined for two jets j1 and j2 as: 
\begin{equation}
\at \equiv \frac{E{\rm _{T}^{j2}}}{M{\rm _{T}^{j1,j2}}},
\label{eq:atone}\end{equation}
where $M{\rm _{T}^{j1,j2}}$ is the transverse mass of the two jets, i.e.:
\begin{equation}
\at \equiv \frac{\sqrt{E{\rm _{T}^{j2}} / E{\rm _{T}^{j1}}}}{\sqrt{2(1-\cos\Delta\phi)}},
\label{eq:attwo}\end{equation}
with $\Delta\phi$ the angle between the two jets. The method is roughly based on the concept of reversing a cut on a control variable ($H_{\rm T}\equiv\sum_{\text{jets}\,j}p_{{\rm T}j}$) to check the (ideally) signal-free region of a discriminating variable (\at). The MC-versus-data agreement in a control sample ($80<H_{\rm T}<120~\gev$) far from the signal region ($H_{\rm T}<350~\gev$) is illustrated in fig.~\ref{fg:cms-at}. A study of the $\at > 0.55$ rejection power as the $H_{\rm T}$ threshold increases demonstrated the robust behaviour of the \at\ cut~\cite{cms-at}. A detailed account of recent SUSY studies performed at CMS with LHC early data is presented in ref.~\cite{widl}, whereas a search for long-lived (`stopped') gluinos is given in ref.~\cite{azzurri} among other searches for New Physics. 

The first LHC collision data at 7~\tev\ collected by ATLAS and CMS also allowed testing the robustness of the \met-based strategy for discovering SUSY. This is illustrated in fig.~\ref{fg:atlas-met}, where the Monte Carlo \met\ distributions for various SM processes that may fake a SUSY-like signal are drawn. The estimates are predominantly based on Monte Carlo simulations, however in some cases these have been scaled to data-deduced factors. The data agree well with the simulation in the 0-lepton plus three-jet selection~\cite{atlas-met}. In the same plot, the prediction for the (low mass, high cross section) mSUGRA point SU4\footnote{SU4: \mz=200~\gev, \mh=160~\gev, $\tan\beta=10$, $A_{0}=-400~\gev$, $\mu>0$.} is superposed. The characteristic long, high-\met\ tail of the production of DM particles discussed in section~\ref{sc:lhc} is clearly visible. More insight on current DM searches performed by ATLAS, either related to supersymmetry or to universal extra dimensions, can be found in ref.~\cite{siragusa}. 

The LHC data processed and analyzed so far are $\mathcal{O}(100~\inb)$, not sufficient to supersede the exclusion limits set by LEP and Tevatron. Nevertheless with the $\sim35~\ipb$ already recorded by ATLAS and CMS and the few inverse femtobarns expected to be provided in the next 1--2~years, a discovery will be within reach. This is demonstrated in fig.~\ref{fg:cms-reach} for the CMS~\cite{cms-2010-008} and in fig.~\ref{fg:atlas-reach} for ATLAS~\cite{atl-phys-pub-2010-010} for integrated luminosities from 100~\ipb\ to 1~\ifb. The largest coverage is achieved through the all-hadronic channel, while comparable reach is provided by the 1-lepton mode. Other channels, e.g.\ the 2-lepton, apart from cross checks, will play a central role ---in the event of a discovery--- in constraining sparticles masses, as discussed in section~\ref{sc:dmlhc}.

\begin{figure}[ht]
\includegraphics[width=0.7\textwidth]{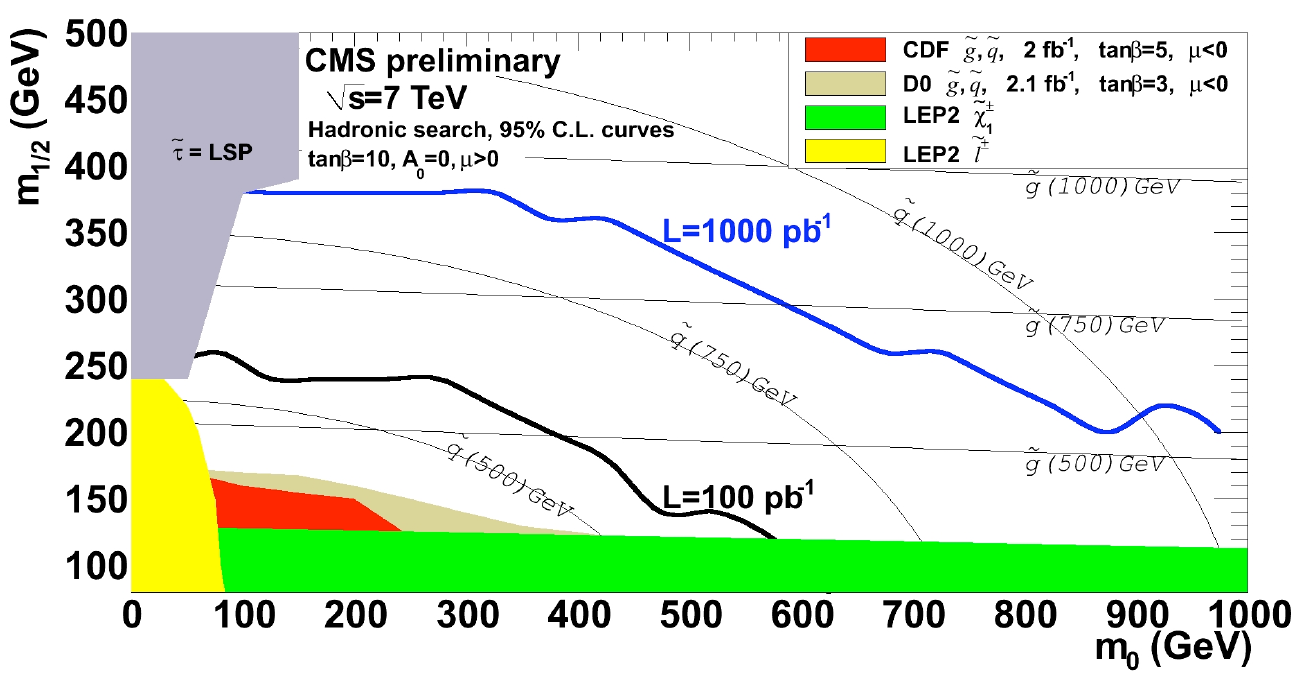}%
\caption{\label{fg:cms-reach} Estimated 95\% C.L.\ exclusion limits for the all-hadronic SUSY search, based on simulated events, expressed in mSUGRA parameter space~\cite{cms-2010-008}.}
\end{figure}

\begin{figure}[ht]
\includegraphics[width=0.45\textwidth]{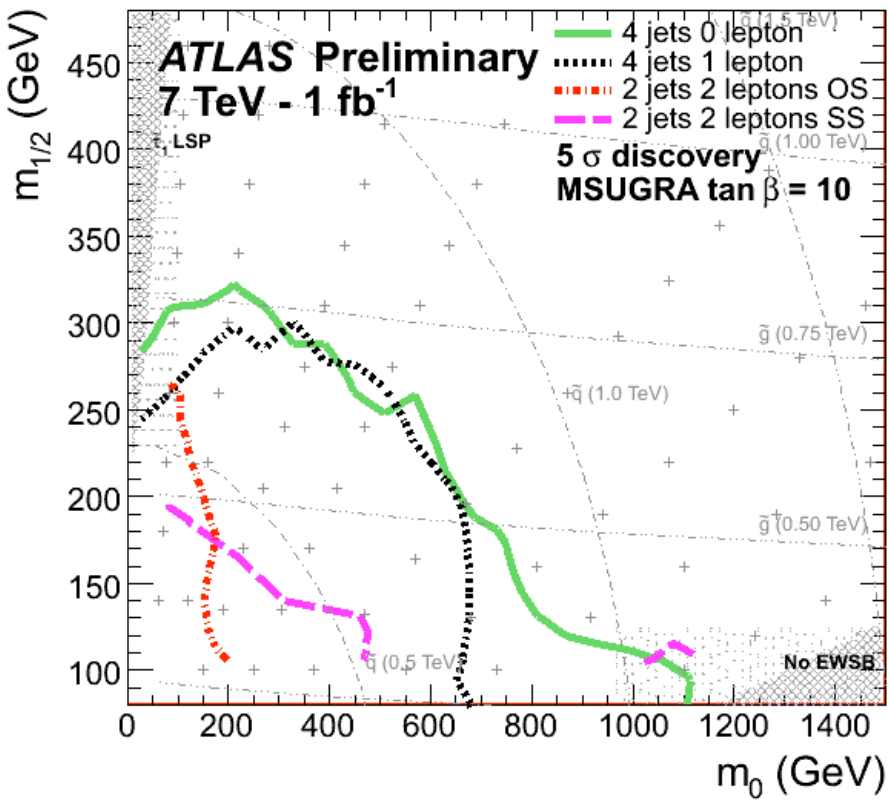}\hspace{2pc}%
\begin{minipage}[b]{0.45\textwidth}\caption{\label{fg:atlas-reach} $5\sigma$ discovery reach as a function of \mz\ and \mh\ for a $\tan\beta=10$ mSUGRA scan for channels with 0,~1 and~2 leptons. The assumed integrated luminosity is 1~\ifb~\cite{atl-phys-pub-2010-010}.\\ }
\end{minipage}
\end{figure}

\section{Pinning down dark matter at LHC}\label{sc:dmlhc}

Once clear evidence for a possible supersymmetric signal is established, the question of whether this implies the existence of SUSY or one of its lookalike arises. In addition, if we assume that it is a SUSY signal, we need to pin down the exact SUSY breaking mechanism and measure its theoretical parameters. Both issues are discussed here, giving some examples of sparticle mass measurements and spin and parameter determination.

\subsection{Constraining sparticle masses}\label{sc:masses}

As an example of the methods aiming at constraining sparticle-mass relations in sparticle cascade decays, we present here a dilepton analysis studied at CMS with simulated data~\cite{cms-dilepton-one,cms-dilepton-two}. Similar analyses with ATLAS are documented in refs.~\cite{atlas-phys-tdr,atlas-ued}. The analysis is targeting an integrated luminosity of 200--300~\ipb\ at $\sqrt{s}=10~\tev$ and its objectives are: (i) to observe a significant excess of opposite-sign same-flavour leptons over the various backgrounds, and (ii) to measure the endpoint in the invariant mass distribution. The latter is directly related to sparticle-masses differences, being sensitive to opposite-sign same-flavour (OS-SF) dileptons coming from the last stages of the decay chain of sparticles: 
\begin{equation}
\t{q} \to \t{\chi}_{2}^{0} \, q \to \t{\ell}^{\pm} \ell^{\mp} q \to \X\,\ell^{\pm} 	\ell^{\mp} q, \qquad \ell = e \text{ or } \mu
\label{eq:dilepton}\end{equation}
The shape of the distribution largely depends on the exact decay chain, so various mSUGRA benchmark points have been considered. For instance at LM0,\footnote{LM0: \mz=200~\gev, \mh=160~\gev, $\tan\beta=10$, $A_{0}=-400~\gev$, $\mu>0$.} the mass difference of the two lightest neutralinos is smaller than the \Z\ boson mass and any slepton mass. Two opposite-sign same-flavour leptons come from the decay chain $\t{\chi}_2^0\to\X\ell^{\pm}\ell^{\mp}$, hence the edge position represents this mass difference: 
\begin{equation}
m_{\ell\ell,{\rm max}} = m_{\t{\chi}_2^0} - m_{\X}.
\label{eq:dilepton-diff}\end{equation}

At the LM1,\footnote{LM1: \mz=60~\gev, \mh=250~\gev, $\tan\beta=10$, $A_{0}=0$, $\mu>0$.} on the other hand, the mass difference of the two lightest neutralinos is larger than the mass of the lightest slepton, so a slepton can be an intermediate product in the neutralino decay chain $\t{\chi}_2^0\to\t{\ell}_R\ell\to\X\ell^{\pm}\ell^{\mp}$. The equation connecting the position of the edge with sparticles masses takes the form: 
\begin{equation}
(m_{\ell\ell,{\rm max}})^2 = \frac{(m_{\t{\chi}_2^0}^{2} - m_{\t{\ell}}^{2})(m_{\t{\ell}}^{2} - m_{\X}^{2})}{m_{\t{\ell}}^{2}}.
\label{eq:dilepton-differ}\end{equation}

The main sources of physics background are uncorrelated supersymmetric decays and SM processes: \ttb, dibosons, and associated production of \W/\Z\ bosons and jets. A data-driven strategy has been developed to eliminate these background processes, based on different-flavour dilepton ($e\mu$) pairs in order to estimate the background in the $ee$ and $\mu\mu$ combinations. It has been demonstrated that the background estimate is reliable for flavour-symmetric background processes.

The position of the endpoint in the invariant mass distribution of the two leptons is then extracted via a maximum-likelihood fit with components describing the background flavour-symmetric processes (see fig.~\ref{fg:cms-edge-1}, left) and the characteristic triangular shape of the supersymmetric signal (see fig.~\ref{fg:cms-edge-1}, right). The latter may be fitted by a function corresponding to a two- or a three-body decay.

\begin{figure}[ht]
\includegraphics[angle=90,width=0.49\textwidth]{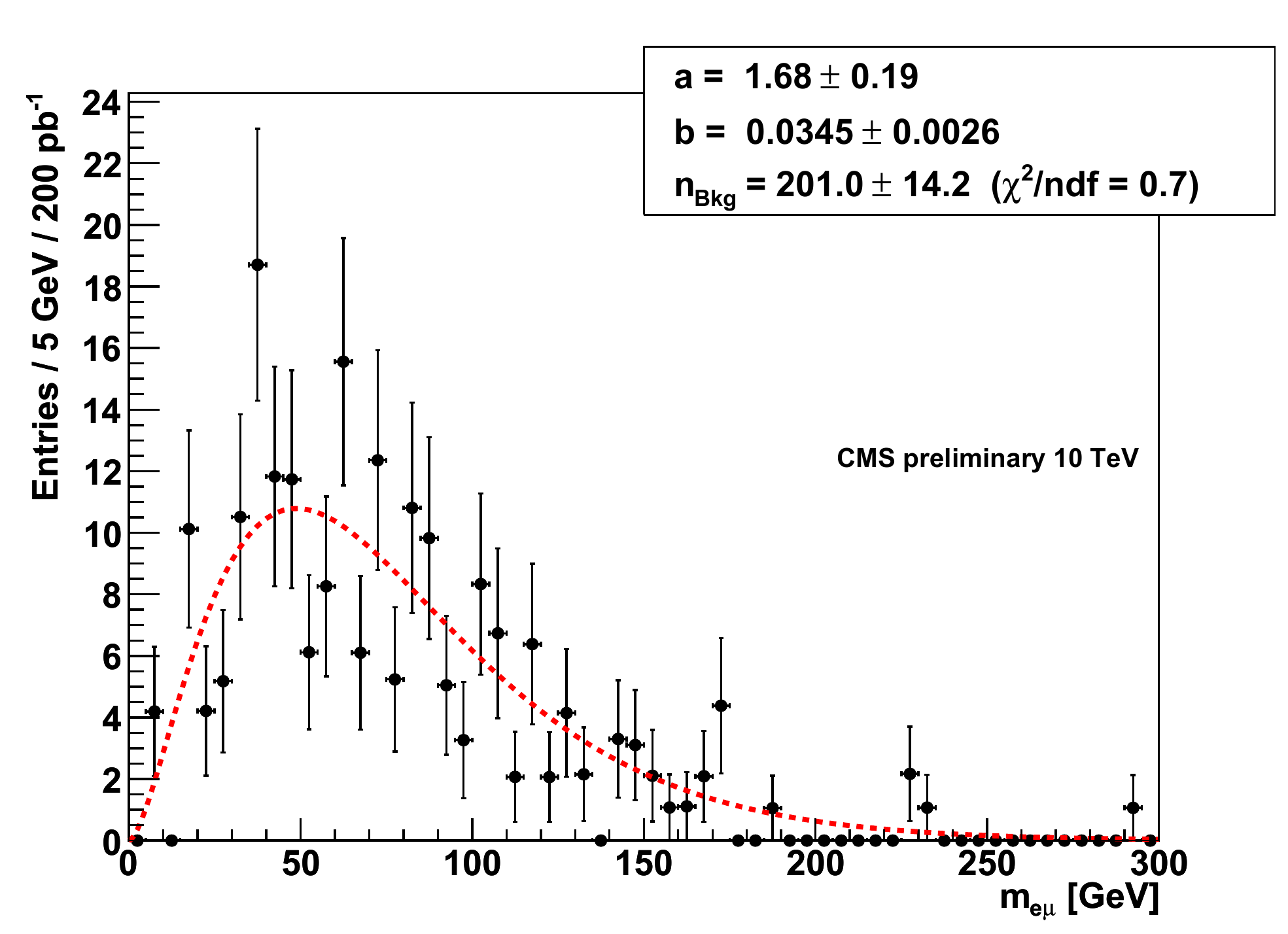}\hfill%
\includegraphics[angle=90,width=0.49\textwidth]{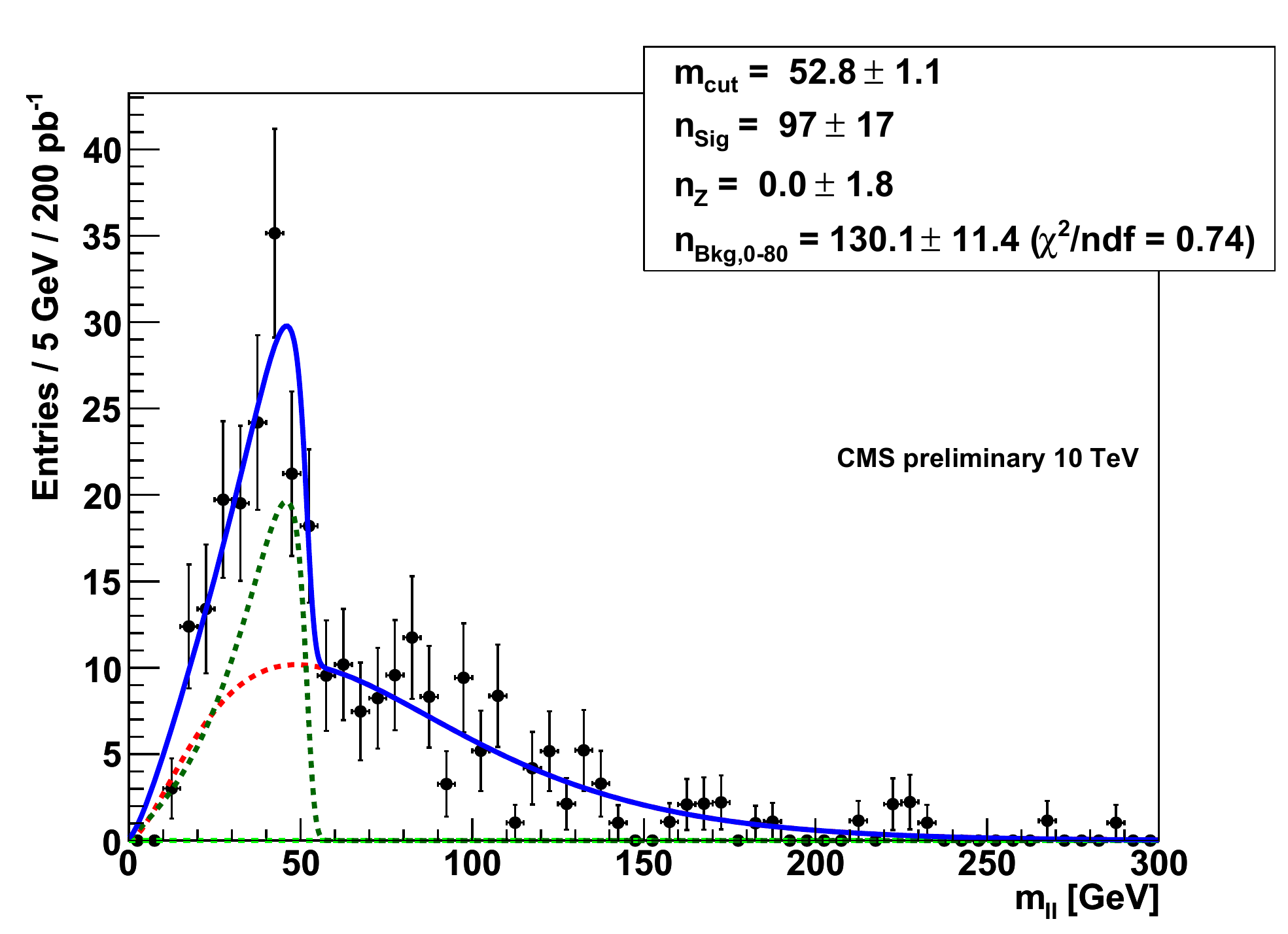}
\caption{\emph{Left:} The fit of the background function to the $e\mu$ invariant mass distribution~\cite{cms-dilepton-two}. \emph{Right:} The combined fit at LM0 for 200~\ipb. The green curve represents the SUSY signal model, the red curve is the background function and the light green dashed line the \Z\ contribution. The black points represent the MC events~\cite{cms-dilepton-two}. } \label{fg:cms-edge-1}
\end{figure}

The number of signal events derived from the fit agrees with the true number of signal events~\cite{cms-dilepton-two}. The theoretical endpoint value is reproduced in case of the fit with the three-body decay model, while the theoretical value is underestimated if the model is fitted for a two-body decay. At the benchmark points LM1 and LM9,\footnote{LM9: \mz=1450~\gev, \mh=175~\gev, $\tan\beta=50$, $A_{0}=0$, $\mu>0$.} a higher integrated luminosity is necessary to measure the endpoint. In other studied benchmark points, an integrated luminosity needed to obtain a $5\sigma$ discovery using shape information of 250~\ipb\ (LM1) and 350~\ipb\ (LM9)~\cite{cms-dilepton-two}. 

Further constraints can be set in more complex combinations of sparticle masses if jets are added to the invariant mass calculation~\cite{atlas-csc}. If sufficient endpoints are known, then the masses themselves can be deduced, e.g.\ by using a numerical $\chi^2$ minimization based on the MINUIT package~\cite{minuit} to extract the SUSY particle masses from a combination of endpoints. A first look at sparticle masses is possible with early data, although with large uncertainties. Appropriate model assumptions and additional information will probably have to be used to constrain the fits.

\subsection{Parameter determination}

The next step after discovery will be to select specific supersymmetric decay chains to measure the properties of the new particles. Here we focus on how a selected set of early studies can be combined to obtain the first measurements of supersymmetric masses and of the parameters of the mSUGRA model with 1~\ifb\ of ATLAS~\cite{atlas-csc} at a centre-of-mass LHC energy of 10~\tev. Specific benchmarks in parameter space have been used to demonstrate the precision that can be expected from these measurements (such as the SU3\footnote{SU3: \mz=100~\gev, \mh=300~\gev, $\tan\beta=6$, $A_{0}=-300~\gev$, $\mu>0$.} point here), but the same (or similar) techniques can be applied to a considerable portion of the SUSY parameter space accessible with LHC data.

A first glimpse of the possible parameter space can be obtained by performing a Markov-chain analysis. With this technique it is possible to efficiently explore a large-dimensional parameter spaces and check whether there are several topologically disconnected parameter regions which are favoured by a given set of measurements. 

A stringent constraint on the SUSY model can be achieved by fitting theoretical calculations for a given set of parameters ---performed by spectrum calculators like SPheno~\cite{spheno}, SoftSUSY~\cite{softsusy} or the ISASUSY~\cite{isasusy} decay package of ISAJET~\cite{isajet}--- to the mass combinations acquired by measuring edge points in invariant distributions. This fitting can be performed by specialized parameter-fitting packages, such as Fittino~\cite{fittino} or SFitter~\cite{sfitter}. In order to estimate the expected precision for such measurements, a number of toy fits for a fixed ${\rm sgn}\,\mu$ has been performed by ATLAS~\cite{atlas-csc}. The four-dimensional distribution of parameters obtained from these toy fits is used to derive the parameter uncertainties and their correlations. The mean values and uncertainties of the results for the parameters $m_0$, $m_{1/2}$, $\tan\beta$ and $A_0$ are listed in table~\ref{tb:parameters}. The parameters $m_0$ and $m_{1/2}$ can be derived reliably with uncertainties of $\mathcal{O}(10~\gev)$, whereas for $\tan\beta$ and $A_0$ only the order of magnitude can be derived from these measurements. The $\chi^2$ distribution of the fits can be used to evaluate the toy-fit performance. The observed mean $\chi^2= 12.6\pm0.2$ for ${\rm sgn}\,\mu=+1$ is compatible with the expected value of $N_{\rm dof} = 11$. The solutions for the wrong assumption ${\rm sgn}\,\mu=-1$, also reported in table~\ref{tb:parameters}, cannot however be ruled out as the observed mean $\chi^2=15.4\pm0.3$ is also acceptable. 

\begin{table}[ht]
  \caption{Results of a fit of the mSUGRA parameters to mass endpoints for the SU3 point in ATLAS~\cite{atlas-csc}. The mean and RMS of the distribution of the results from toy fits is reported. The two possible assumptions for the digital parameter ${\rm sgn}\,\mu=+1$, ${\rm sgn}\,\mu=\pm1$ have been used, resulting in different preferred regions for the other parameters. The experimental uncertainties are also shown.}\label{tb:parameters}
  \begin{center}
  \lineup
  \begin{tabular}{*{4}{l}}                                     \br
Parameter & SU3 value & Fitted value & Uncertainty \\  \mr
     \multicolumn{4}{c}{${\rm sgn}\,\mu=+1$} \\ \mr
$\tan\beta$ & \0\06 & \0\07.4 & \0\04.6 \\
$m_0$ [\gev] & 100 & \098.5 & \0\09.3 \\
$m_{1/2}$ [\gev] & 300 & 317.7 & \0\06.9 \\
$A_0$ [\gev] & \-300 & 445 & 408 \\  \mr
    \multicolumn{4}{c}{${\rm sgn}\,\mu=-1$} \\ \mr
$\tan\beta$ &   & \013.9 & \0\02.8 \\
$m_0$ [\gev] &   & 104 & \018 \\
$m_{1/2}$ [\gev] &   & 309.6 & \0\05.9 \\
$A_0$ [\gev] &   & 489 & 189 \\  \br  \end{tabular}
\end{center}
\end{table}

Hence with 1~\ifb\ the reconstruction of part of the supersymmetric mass spectrum will only be possible for favourable SUSY scenarios and with some assumptions about the decay chains involved. Larger integrated luminosity will help to overcome these limitations, as more measurements become possible and the precision of each increases. Furthermore the mass spectrum constraints in conjunction with precision observables, such as the $(g-2)_{\mu}$ and the $b\to s\gamma$, will illuminate the flavour mixing and possibly the $CP$ properties of the supersymmetric model~\cite{neil}.

\subsection{Spin measurement}

Measurements of the number of new particles and their masses will provide us enough information to extract model parameters for one of the SUSY models. However, the mass information alone will not be enough to distinguish different new physics scenarios. For example universal extra dimensions~\cite{ued} with Kaluza-Klein parity can have a mass spectrum very similar to the one of certain SUSY models. However, the spin of the new particles is different and can be used to discriminate between models~\cite{Smillie:2005ar}. Another method based on robust ratios of inclusive counts of simple physics objects has also been proposed~\cite{Hubisz:2008gg}.

In order to measure the spin of newly discovered particle, one possibility is to use two-body slepton decay chains as the ones described earlier in this section. In particular the cascade decay of the $\t{q}_L$ to $\t{\chi}_2^0$ which further decays to slepton (fig.~\ref{fg:atl-spin-cascade}) is very convenient for such measurements~\cite{Barr:2004ze}.
\begin{figure}[ht]
\includegraphics[width=0.35\textwidth]{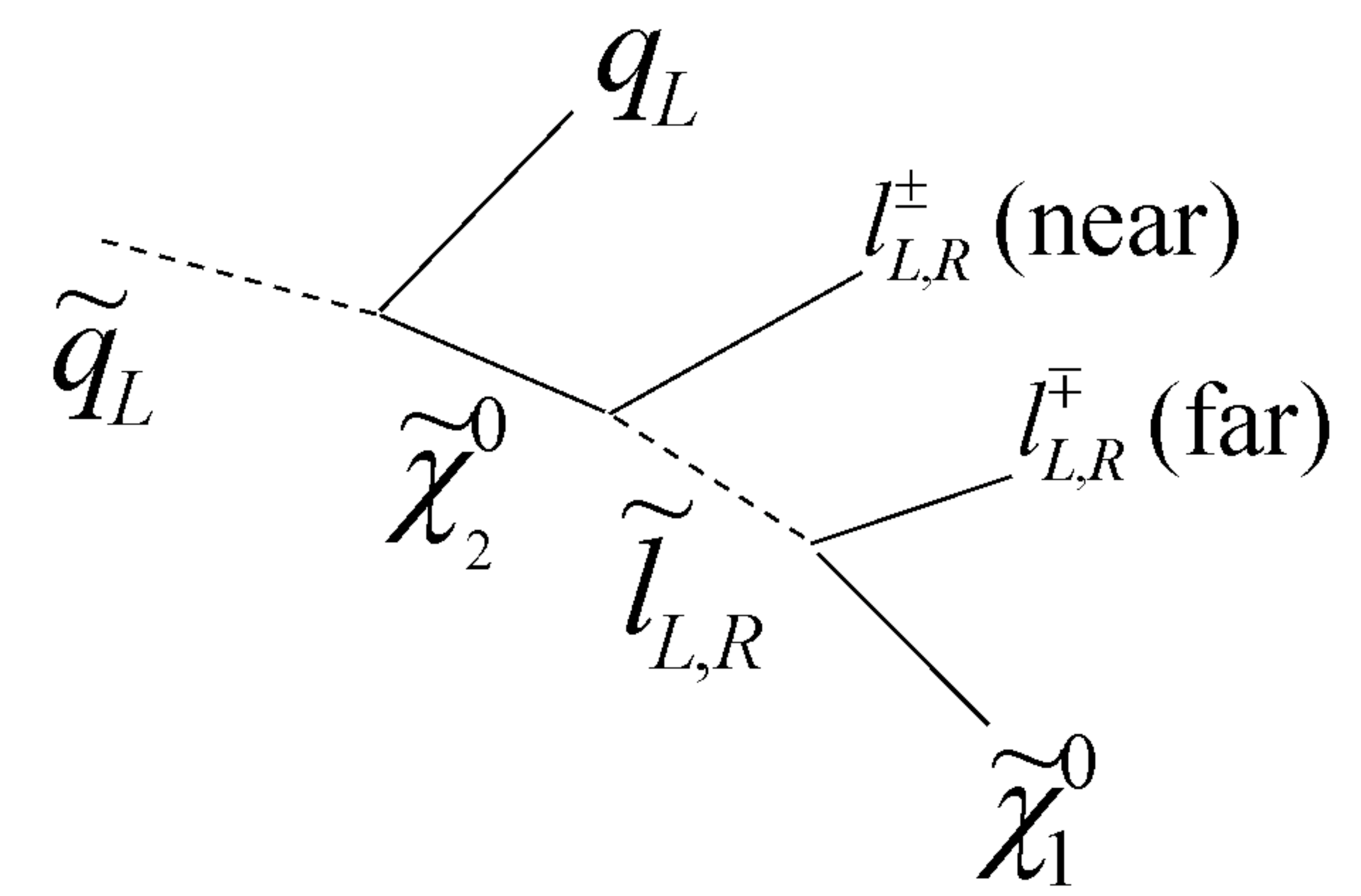}\hspace{2pc}%
\begin{minipage}[b]{0.5\textwidth}\caption{\label{fg:atl-spin-cascade} Schematic view of $L$-type squark decay. The lepton from the $\t{\chi}_2^0$ decay is called \emph{near}; the lepton from $\t{\ell}_{L,R}$ decay is called \emph{far}.}
\end{minipage}
\end{figure}

The charge asymmetry of $\ell q$ pairs, for instance, can be used to measure the spin of $\t{\chi}_2^0$, while the shape of dilepton invariant mass spectrum measures slepton spin~\cite{Biglietti:2007mj}. The first lepton in the decay chain is called the \emph{near} lepton while the other is called the \emph{far} lepton. The invariant masses $m(q\ell_{\rm near})$ charge asymmetry $A$ is trivially defined as:
\begin{equation}\label{eq:asymmetry}
A \equiv \frac{s^+-s^-}{s^++s^-}\, ,
\end{equation}
where $s^{\pm} = {\rm d}\sigma/{\rm d}m(q\ell_{\rm near}^{\pm})$.

In general is not possible to distinguish between the near and the far lepton and only $m(\bar{q}\ell_{\rm near})$ can be measured, diluting $A$. The expected asymmetry for the mSUGRA benchmark point SU3, as estimated by ATLAS for $\sqrt{s}=14~\tev$, is shown in fig.~\ref{fg:atl-spin} for a luminosity of 30~\ifb~\cite{Biglietti:2007mj}.

\begin{figure}[ht]
\includegraphics[width=0.48\textwidth]{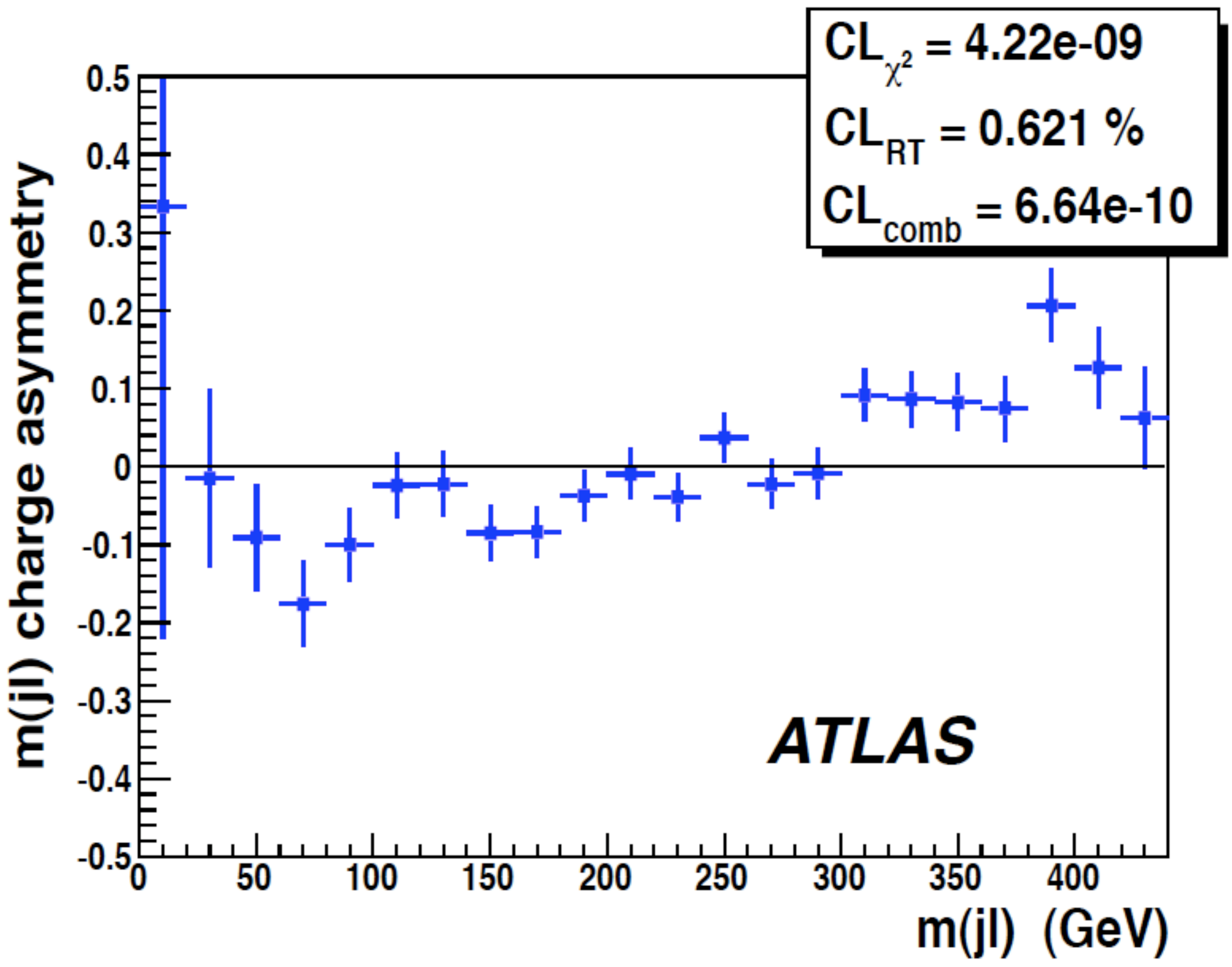}\hspace{2pc}%
\begin{minipage}[b]{0.45\textwidth}\caption{\label{fg:atl-spin}Charge asymmetry for lepton-jet invariant mass after SFOS-OFOS subtraction using both near and far leptons in SU3 point~\cite{Biglietti:2007mj}.\\ }
\end{minipage}
\end{figure}

Results show that, in a fast simulation approach without taking into account systematic effects coming from a realistic detector description, an integrated luminosity of at least 100~\ifb\ is needed in the case of the SU1\footnote{SU1: \mz=70~\gev, \mh=350~\gev, $\tan\beta=10$, $A_{0}=0$, $\mu>0$.} point to observe a non-zero charge asymmetry with a confidence level of about 99\%, while in the more favourable case of the SU3 point 10~\ifb\ would be sufficient~\cite{Biglietti:2007mj}. It becomes therefore evident that even if the LHC experiments observe a SUSY-like signal during the next two years (2011--2012) of operation at the `low' LHC energies of $7-8~\tev$, much more data at higher energy of $14~\tev$ will be necessary to establish the identity of the underlying theory.

A significant role in this context will be played by the high precision measurements expected to be performed at the ILC~\cite{ilc,dm-colliders}. If the determination of the properties of the DM particle by collider experiments~\cite{dm-colliders} matches cosmological observations to high precision, then (and only then) we will be able to claim to have determined what DM is. Such an achievement would be a great success of the Particle-Physics/Cosmology connection and would give us confidence in our understanding of the Universe. 

\section{Interplay between the dark sector and LHC: Alternative scenarios}\label{sc:alt}

The simplest proposal to explain the origin of dark energy is to add a cosmological constant to Einstein's equation~\cite{lcdm}. However, the reason why the dark matter content is  comparable to the dark energy content at the present time remains a puzzle. Modifications to general relativity, braneworld scenarios, and topological defects are some of the proposals attempting to explain this fundamental issue. In string theory, the dilaton can play the role of dark energy~\cite{dm}. In this section we review experimental signatures of SUSY as consequences of a rolling dilaton in the Q-cosmology scenario~\cite{dm,mm1} which offers an alternative framework that establishes the Supercritical (or non-critical) String Cosmology (or SSC)~\cite{dm}. Such a dilaton modifies the Boltzmann equation, thus diluting the supersymmetric dark matter density (of neutralinos) by a factor $\mathcal{O}(10)$~\cite{dm-modified,lahanas} and consequently the parameter-space regions excluded by the standard scenario are allowed in the SSC. Such deviating predictions in the DM relic abundance also arise in the context of space-time (D-particle) foam in string/brane-theories~\cite{mavro}. However in such cases, the effects of the D-particle foam on the relic abundance are opposite to those of the dilaton in the SSC models, but their magnitude depends on the string scale and thus such effects can be relevant to LHC only for low (\tev) string scales.

The mSUGRA final states at the LHC favoured by supercritical string cosmology have been studied in depth in ref.~\cite{dutta}. It becomes evident by inspecting the two panels in fig.~\ref{fg:dutta}, that the dark-matter-allowed region has larger values of \mz\ compared to the standard cosmology case. Thus, the final states in the SSC scenario are different from those of the standard cosmology. For example, in the case of standard cosmology for smaller values of \mz\ (also allowed by the $(g-2)_{\mu}$ constraint), we have low-energy taus in the final state due to the proximity of the stau to the neutralino mass in the stau-neutralino coannihilation region. On the other hand, in the SSC case the final states contain \Z\ bosons, Higgs bosons or high-energy taus. 

\begin{figure}[ht]
\includegraphics[width=0.48\textwidth]{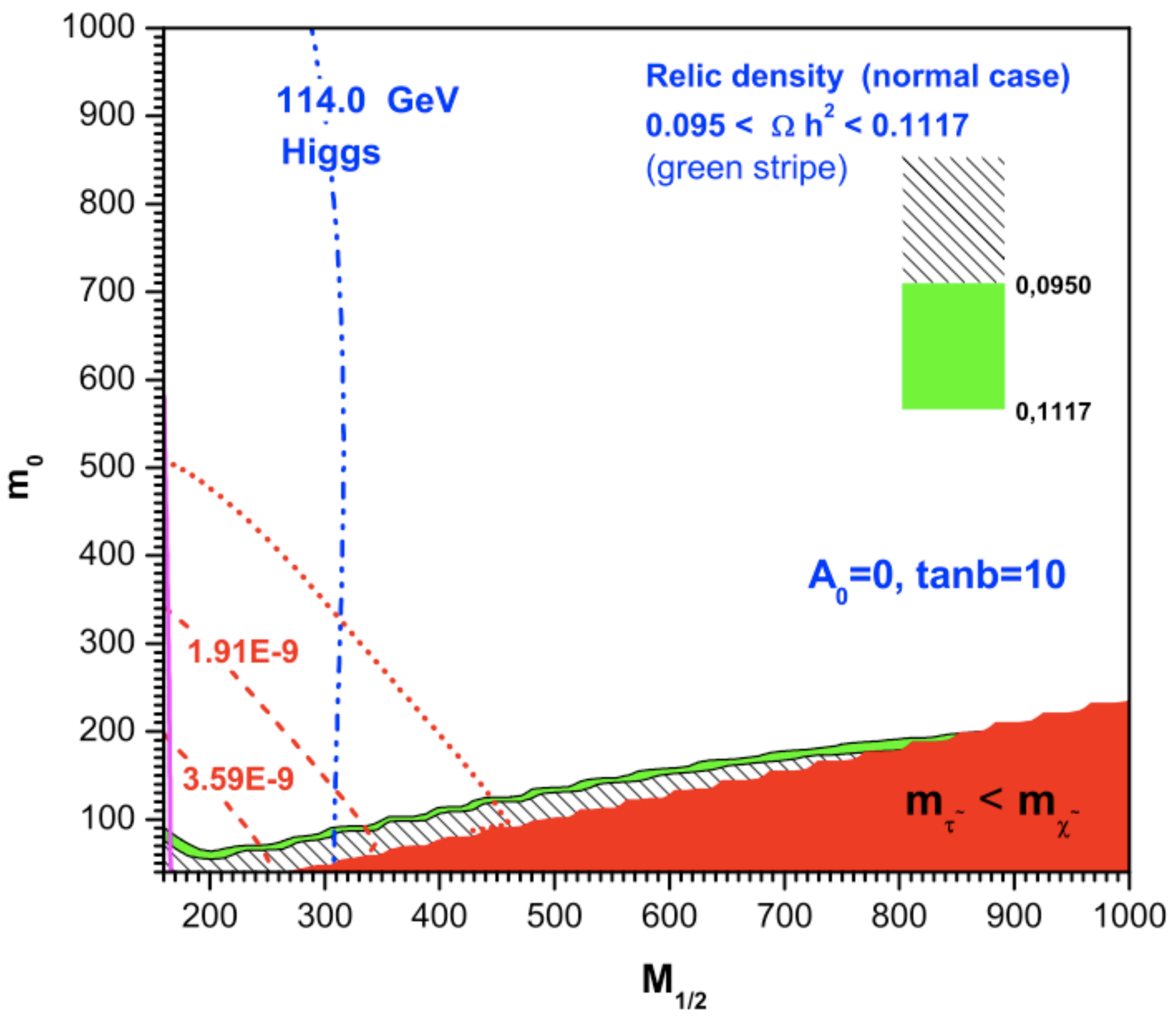}\hspace{1pc}
\includegraphics[width=0.48\textwidth]{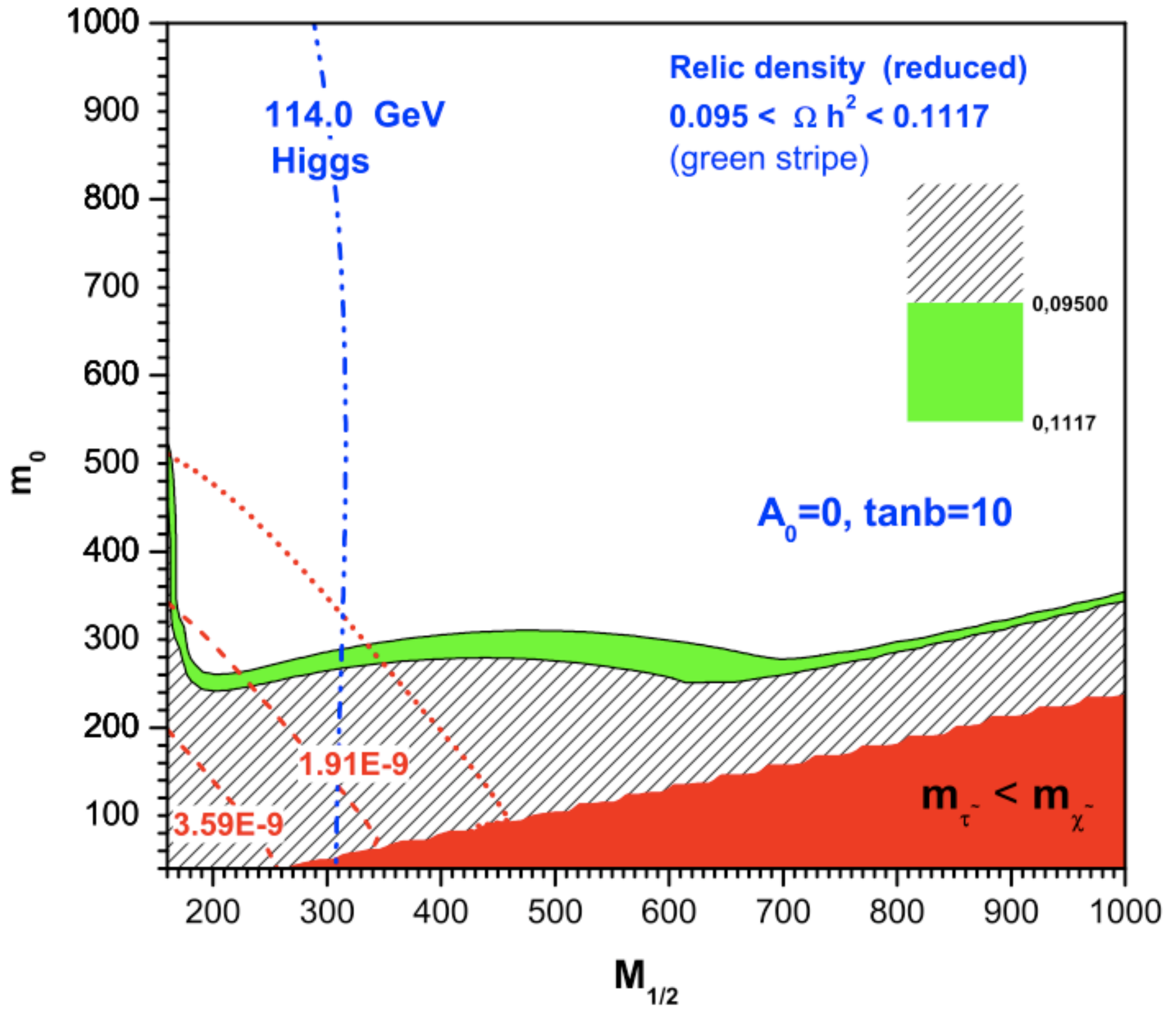}
\caption{\label{fg:dutta} WMAP3-allowed parameter space in mSUGRA $(\mz,\,\mh)$ plane for the standard cosmology (left) and the SSC (right) for $A_0=0$ and $\tan\beta=10$: regions where the neutralino relic density is within the WMAP3 limits (green stripe) and where it is lower than this (hatched region) are shown. Also shown are the $h^0$ mass boundary (dash-dotted blue), the $(g-2)_{\mu}$ $1\sigma$ (dashed red) and $2\sigma$ (dotted red) boundaries and the stau-LSP region (lower solid red)~\cite{dm-modified}.}
\end{figure}

In fact these final states dominate in most of the allowed mSUGRA parameter space. Therefore, by analyzing the parameter space of the SSC model, most regions of the mSUGRA parameter space at the LHC are explored. The following final states have been studied~\cite{dutta}: 

\vspace*{-0.1cm}
\begin{table*}[htdp]
\begin{center}
\begin{tabular}{rl}
Higgs decays: & ($h^{0}\to$) $b\bar{b}$ + jets + \met \\
\Z\ boson decays: &   ($Z\to$) $\ell^{\pm}\ell^{\mp}$ + jets + \met \\
Ditau channel: &   $2\tau$  + jets + \met
\end{tabular}
\end{center}
\end{table*}
\vspace*{-0.5cm}

\noindent as well as constructed observables such as the endpoints of invariant mass distributions $M_{bbj}$, $M_{\ell\ell j}$, and $M_{\tau\tau j}$ and the peak position of $M_{\tau\tau}$. All of these analyses have been studied with ATLAS and/or CMS with simulated data at $\sqrt{s}=10\;\text{or}\;14~\tev$: $h^{0}\to b\bar{b}$ channel~\cite{atlas-phys-tdr,thesis,atlas-csc,cms-tdr}, the $Z\to\ell^{\pm}\ell^{\mp}$ mode~\cite{cms-tdr} and the 2$\tau$ mode~\cite{atlas-csc,cms-tdr}. In the future, when $pp$ collision data at an energy of 14~\tev\ will be accumulated, searches for SUSY with these signatures will be possible.  

It is remarked that the SSC scenario is consistent with the smoothly evolving dark energy at least for $0<z<1.6$, in accordance with the observations on supernovae~\cite{mm1}, on the galaxy-ages-measured Hubble rate~\cite{mm2} and on the baryon acoustic oscillations~\cite{mm3}. Hence it offers a cosmologically viable solution.

\section{Outlook}\label{sc:out}

The origin of dark matter remains one of the most compelling mysteries in our understanding of the Universe today and the Large Hadron Collider, already delivering $pp$ collision data at CERN at an unprecedented high-energy, is going to play a central role in constraining some of its parameters. A deviation from SM in inclusive signatures like missing energy plus jets (plus leptons) will hint a discovery of DM, however exclusive studies are required to roughly determine the new-particle properties and model parameters. Although the scheme is developed with SUSY in mind, it is applicable to other beyond-standard-model scenarios such as UED and $T$-parity Little Higgs.

If LHC should discover general WIMP dark matter, it will be non-trivial to prove that it has the right properties. Future $e^+e^-$ colliders (ILC, CLIC) are expected to extend the LHC discovery potential and improve the identification of the underlying DM model. By providing more precise determination of model parameters, they will consequently put bounds on relic density, direct detection rate and WIMP annihilation processes.

The complementarity between LHC and cosmo/astroparticle experiments lies in the uncorrelated systematics and the measurement of different model parameters. In the following years we expect a continuous interplay between particle physics experiments and astrophysical/cosmological observations.

\ack
The author is grateful to the DISCRETE2010 organizers and especially Antonio Di~Domenico for the kind invitation and support. Thanks to them, a warm, friendly and intellectually stimulating atmosphere was enjoyed by the speakers and participants throughout the Symposium. This work was supported in part by the Spanish Ministry of Science and Innovation (MICINN) under the project FPA2009-13234-C04-01 and by the Spanish Agency of International Cooperation for Development under the PCI projects A/023372/09 and A/030322/10. The author acknowledges support by the CERN Corresponding Associate Programme.



\section*{References}
\begin{thebibliography}{99}

\bibitem{dm-review} For a pedagogical introduction, see e.g.: Hooper~D 2009 TASI 2008 Lectures on Dark Matter {\it Preprint} arXiv:0901.4090 [hep-ph]

\bibitem{lhc} Evans~L and Bryant~P 2008 LHC Machine {\it J.\ Instrum.} {\bf 3} S08001

\bibitem{direct} Akimov~D 2011 Techniques and Results for the Direct Detection of Dark Matter (Review) {\it Nucl.\ Instrum.\ Meth.} A {\bf 628} 50--58 {\it and references therein}

\bibitem{indirect} Morselli~A 2011 Indirect detection of dark matter, current status and recent results {\it Prog.\ Part.\ Nucl.\ Phys.} {\bf 66} 208--215 {\it and references therein} 

\bibitem{ilc} Battaglia~M 2009 The role of an $e^+e^-$ linear collider in the study of cosmic dark matter {\it New J.\ Phys.} {\bf 11} 105025 \\
Aarons~G {\it et al} [ILC Collaboration] 2007 International Linear Collider Reference Design Report Volume~2: Physics at the ILC {\it Preprint} arXiv:0709.1893 [hep-ph]

\bibitem{clic} Accomando~E {\it et al}  [CLIC Physics Working Group] Physics at the CLIC multi-TeV linear collider {\it Preprint} arXiv:hep-ph/0412251


\bibitem{mond} For an review on MOND, see e.g.: Sanders~R~H and McGaugh~S~S 2002 Modified Newtonian Dynamics as an Alternative to Dark Matter {\it Ann.\ Rev.\ Astron.\ Astrophys.} {\bf 40} 263--317 ({\it Preprint} arXiv:astro-ph/0204521)

\bibitem{khlopov} Glashow~S~L 2005 A Sinister extension of the standard model to $SU(3)\times SU(2)\times SU(2)\times U(1)$ {\it Preprint} arXiv:hep-ph/0504287 \\
Khlopov~M~Y and Kouvaris~C 2008 Strong Interactive Massive Particles from a Strong Coupled Theory {\it Phys.\ Rev.} D {\bf 77} 065002 ({\it Preprint} arXiv:0710.2189 [astro-ph]) {\it and references therein}



\bibitem{snIa} Amanullah~R {\it et al} 2010 Spectra and Light Curves of Six Type~Ia Supernovae at $0.511<z<1.12$ and the Union2 Compilation {\it Astrophys.\ J.} {\bf 716} 712--738 ({\it Preprint} arXiv:1004.1711 [astro-ph.CO])

\bibitem{wmap} Komatsu~E {\it et al} 2011 Seven-Year Wilkinson Microwave Anisotropy Probe (WMAP) Observations: Cosmological Interpretation {\it Astrophys.\ J.\ Suppl.} {\bf 192} ({\it Preprint} arXiv:1001.4538 [astro-ph.CO])

\bibitem{bao} Eisenstein~D~J {\it et al} [SDSS Collaboration] 2005 Detection of the Baryon Acoustic Peak in the Large-Scale Correlation Function of SDSS Luminous Red Galaxies {\it Astrophys.\ J.} {\bf 633} 560--574 ({\it Preprint} astro-ph/0501171) {\it and references therein}

\bibitem{lensing} Broadhurst~T, Umetsu~K, Medezinski~E, Oguri~M and Rephaeli~Y 2008 Comparison of Cluster Lensing Profiles with Lambda CDM Predictions {\it Astrophys.\ J.} {\bf 685} L9--L12 ({\it Preprint} arXiv:0805.2617 [astro-ph]) {\it and references therein}

\bibitem{cmb} Larson~D {\it et al} 2011 Seven-Year Wilkinson Microwave Anisotropy Probe (WMAP) Observations: Power Spectra and WMAP-Derived Parameters {\it Astrophys.\ J.\ Suppl.}  {\bf 192} 16 ({\it Preprint} arXiv:1001.4635 [astro-ph.CO])

\bibitem{lcdm} Carroll~S~M, Press~W~H and Turner~E~L 1992 The Cosmological constant {\it Ann.\ Rev.\ Astron.\ Astrophys.} {\bf 30} 499--542

\bibitem{axion} Zioutas~K, Tsagri~M, Papaevangelou~T, Dafni~T and Anastassopoulos~V 2009 Axion Searches with Helioscopes and astrophysical signatures for axion(-like) particles {\it New J.\ Phys.} {\bf 11} 105020 ({\it Preprint} arXiv:0903.1807 [astro-ph.SR])

\bibitem{shdm} Chung~D~J~H, Kolb~E~W and Riotto~A 1999 Superheavy dark matter {\it Phys.\ Rev.} D {\bf 59} 023501 ({\it Preprint} arXiv:hep-ph/9802238)


\bibitem{susy} Kazakov~D~I 2010 Supersymmetry on the Run: LHC and Dark Matter {\it Nucl.\ Phys.\ Proc.\ Suppl.} {\bf 203-204} 118--154 ({\it Preprint} arXiv:1010.5419 [hep-ph])

\bibitem{ued} Hooper~D and Profumo~S 2007 Dark matter and collider phenomenology of universal extra dimensions {\it Phys.\ Rept.} {\bf 453} 29--115 ({\it Preprint} arXiv:hep-ph/0701197)

\bibitem{little} Birkedal~A, Noble~A, Perelstein~M and Spray~A 2006 Little Higgs dark matter {\it Phys.\ Rev.} D {\bf 74} 035002 ({\it Preprint} arXiv:hep-ph/0603077)
  
\bibitem{grigoris} Panotopoulos~G 2011 The degenerate scenario in the NMSSM: Direct singlino-like neutralino searches with a gravitino LSP {\it Preprint} arXiv:1103.0140 [hep-ph]  {\it in these proceedings}
  
\bibitem{rpv} Barbier~R {\it et al} 2005 R-parity violating supersymmetry {\it Phys.\ Rept.}  {\bf 420} 1--202 ({\it Preprint} arXiv:hep-ph/0406039)

\bibitem{rpv-grav} Takayama~F and Yamaguchi~M 2000 Gravitino dark matter without \R-parity {\it Phys.\ Lett.} B {\bf 485} 388--392 ({\it Preprint} arXiv:hep-ph/0005214) \\
Hirsch~M, Porod~W and Restrepo~D 2005 Collider signals of gravitino dark matter in bilinearly broken \R-parity {\it J. High Energy Phys.} JHEP03(2005)062 ({\it Preprint} arXiv:hep-ph/0503059) \\
Buchmuller~W, Covi~L, Hamaguchi~K, Ibarra~A and Yanagida~T 2007 Gravitino dark matter in \R-parity breaking vacua {\it J. High Energy Phys.} JHEP03(2007)037 ({\it Preprint} arXiv:hep-ph/0702184) 


\bibitem{rpv-axino} Chun~E~J and Kim~H~B 2006 Axino Light Dark Matter and Neutrino Masses with \R-parity Violation {\it J. High Energy Phys.} JHEP10(2006)082 ({\it Preprint} arXiv:hep-ph/0607076)

\bibitem{lola} Lola~S 2009 Gravitino dark matter, neutrino masses and lepton flavor violation from broken \R-parity {\it AIP Conf.\ Proc.} {\bf 1115} 318--323 {\it and references therein} 


\bibitem{atlas-det} Aad~G {\it et al} [ATLAS Collaboration] 2008 The ATLAS Experiment at the CERN Large Hadron Collider {\it J.\ Instrum.} {\bf 3} S08003

\bibitem{cms-det} Adolphi~R {\it et al} [CMS Collaboration] 2008 The CMS experiment at the CERN LHC {\it J.\ Instrum.} {\bf 3} S08004

\bibitem{atlas-csc} Aad~G {\it et al} [ATLAS Collaboration] 2009 Expected Performance of the ATLAS Experiment -- Detector, Trigger and Physics {\it CERN Report} CERN-OPEN-2008-020 {\it Preprint} arXiv:0901.0512 [hep-ex]

\bibitem{cms-tdr} Bayatian G L {\it et al} [CMS Collaboration] 2007 CMS technical design report, volume II: Physics performance {\it J.\ Phys.} G {\bf 34} 995--1579

\bibitem{alan}ÊCourtesy of Alan Barr

\bibitem{geant4} Agostinelli~S {\it et al}  [GEANT4 Collaboration] 2003 GEANT4: A simulation toolkit {\it Nucl.\ Instrum.\ Meth.} A {\bf 506} 250--303 \\
Allison~J {\it et al} 2006 Geant4 developments and applications {\it IEEE Trans.\ Nucl.\ Sci.} {\bf 53} 270--278
  
\bibitem{cms-at} CMS Collaboration 2010 Performance of Methods for Data-Driven Background
Estimation in SUSY Searches {\it CMS Physics Analysis Summary} CMS-PAS-SUS-10-001 

\bibitem{widl} Widl~E 2011 Searches for Supersymmetry with the CMS detector at the LHC {\it in these proceedings}

\bibitem{azzurri}ÊAzzurri~P 2011 First Results of Searches for New Physics at sqrt(s)= 7~TeV with the CMS detector {\it Preprint} arXiv:1103.1048 [hep-ex] {\it in these proceedings}

\bibitem{atlas-met} ATLAS Collaboration 2010 Early supersymmetry searches in channels with jets and missing transverse momentum with the ATLAS detector {\it ATLAS Note} ATLAS-CONF-2010-065

\bibitem{siragusa} Siragusa~G 2011 New Physics with ATLAS: experimental prospects {\it ATLAS Note} ATL-PHYS-PROC-2011-011 {\it in these proceedings} 

\bibitem{cms-2010-008} CMS Collaboration 2010 The CMS physics reach for searches at 7~TeV {\it CMS Note} CMS-NOTE-2010-008

\bibitem{atl-phys-pub-2010-010} ATLAS Collaboration 2010 Prospects for Supersymmetry discovery based on inclusive searches at a 7~TeV centre-of-mass energy with the ATLAS detector {\it ATLAS Note} ATL-PHYS-PUB-2010-010


\bibitem{cms-dilepton-one} CMS Collaboration 2009 Dilepton + Jets + MET channel:
Observation and Measurement of $\t{\chi}_2^0\to\X\ell\ell$ {\it CMS Physics Analysis Summary} CMS-PAS-SUS-08-001

\bibitem{cms-dilepton-two} CMS Collaboration 2009 Discovery potential and measurement of a dilepton mass edge in SUSY events at $\sqrt{s}=10~\tev$ {\it CMS Physics Analysis Summary} CMS-PAS-SUS-09-002

\bibitem{atlas-phys-tdr} Airapetian~A {\it et al} [ATLAS Collaboration] 1999 
  {\it ATLAS: Detector and physics performance technical design report}  vol~1 (Geneva: CERN)  475~p ({\it CERN Report} CERN-LHCC-99-14) \\ 
  Airapetian~A {\it et al} [ATLAS Collaboration] 1999 
  {\it ATLAS: Detector and physics performance technical design report}  vol~2 (Geneva: CERN)  519~p ({\it CERN Report} CERN-LHCC-99-15) 
   
\bibitem{atlas-ued} ATLAS Collaboration 2009 Prospects for Supersymmetry and Univeral Extra Dimensions discovery based on inclusive searches at a 10~\tev\ centre-of-mass energy
with the ATLAS detector {\it ATLAS Note} ATL-PHYS-PUB-2009-084 

\bibitem{minuit} James~F and Roos~M 1975 MINUIT: A System for Function Minimization and Analysis of the Parameter Errors and Correlations {\it Comput.\ Phys.\ Commun.} {\bf 10} 343--367

\bibitem{spheno} Porod~W 2003 SPheno, a program for calculating supersymmetric spectra, SUSY particle decays and SUSY particle production at $e^+e^-$ colliders {\it Comput.\ Phys.\ Commun.}  {\bf 153} 275--315 ({\it Preprint} arXiv:hep-ph/0301101)

\bibitem{softsusy} Allanach~B~C 2002 SOFTSUSY: a program for calculating supersymmetric spectra {\it Comput.\ Phys.\ Commun.} {\bf 143} 305--331 ({\it Preprint} arXiv:hep-ph/0104145)

\bibitem{isasusy} Baer~H, Paige~F~E, Protopopescu~S~D and Tata~X 1993 Simulating Supersymmetry With ISAJET~7.0 / ISASUSY~1.0 {\it Preprint} arXiv:hep-ph/9305342

\bibitem{isajet} Paige~F~E, Protopopescu~S~D, Baer~H and Tata~X 2003 ISAJET~7.69: A Monte Carlo event generator for $pp$, $\bar{p}p$, and $e^+e^-$ reactions {\it Preprint} arXiv:hep-ph/0312045

\bibitem{fittino} Bechtle~P, Desch~K and Wienemann~P 2006 Fittino, a program for determining MSSM parameters from collider observables using an iterative method {\it Comput.\ Phys.\ Commun.} {\bf 174} 47--70 ({\it Preprint} arXiv:hep-ph/0412012)

\bibitem{sfitter} Lafaye~R, Plehn~T and Zerwas~D 2004 SFITTER: SUSY parameter analysis at LHC and LC {\it Preprint} arXiv:hep-ph/0404282
  
\bibitem{neil} Hodgkinson~R~N 2011 Decoding new physics at 1~\ifb\ LHC with flavour and CP observables {\it in these proceedings}

\bibitem{Smillie:2005ar} Smillie~J~M and Webber~B~R 2005 Distinguishing Spins in Supersymmetric and Universal Extra Dimension Models at the Large Hadron Collider
  {\it J. High Energy Phys.}  	JHEP10(2005)069 ({\it Preprint} arXiv:hep-ph/0507170) \\
Datta~A, Kong~K and Matchev~K~T 2005 Discrimination of supersymmetry and universal extra dimensions at hadron colliders {\it Phys.\ Rev.} D {\bf 72} 096006,  
 Erratum-ibid.\ 2005 {\bf 72} 119901 ({\it Preprint} arXiv:hep-ph/0509246)

\bibitem{Hubisz:2008gg} Hubisz~J, Lykken~J, Pierini~M and Spiropulu~M 2008 Missing energy look-alikes with 100~pb$^{-1}$ at the LHC {\it Phys.\ Rev.} D {\bf 78} 075008 ({\it Preprint} arXiv:0805.2398 [hep-ph])

\bibitem{Barr:2004ze} Barr~A~J 2004 Using lepton charge asymmetry to investigate the spin of supersymmetric particles at the LHC {\it Phys.\ Lett.} B {\bf 596} 205--212 ({\it Preprint} arXiv:hep-ph/0405052)

\bibitem{Biglietti:2007mj} Biglietti~M {\it et al} 2007 Study of second lightest neutralino $\t{\chi}_2^0$ spin measurement with ATLAS detector at LHC {\it ATLAS Note} ATL-PHYS-PUB-2007-004

\bibitem{dm-colliders} Baltz~E~A, Battaglia~M, Peskin~M~E and Wizansky~T 2006 Determination of dark matter properties at high-energy colliders {\it Phys.\ Rev.}  D {\bf 74} 103521
({\it Preprint} arXiv:hep-ph/0602187)
  

\bibitem{dm} Mavromatos~N~E 2008 LHC Physics and Cosmology {\it Fundamental Interactions: Proc.\ Lake Louise Winter Institute 2007 (Lake Louise)} ed A Astbury {\it et al} (Singapore: World Scientific) pp~80--127 ({\it Preprint} arXiv:0708.0134 [hep-ph]) 
  
\bibitem{mm1} Ellis~J~R, Mavromatos~N~E, Mitsou~V~A and Nanopoulos~D~V 2007 Confronting dark energy models with astrophysical data {\it Astropart.\ Phys.} {\bf 27} 185--198
({\it Preprint} arXiv:astro-ph/0604272)\\
Mitsou~V~A 2008 Constraints on Dissipative Non-Equilibrium Dark Energy Models from Recent
Supernova Data {\it Fundamental Interactions: Proc.\ Lake Louise Winter Institute 2007 (Lake Louise)} ed A Astbury {\it et al} (Singapore: World Scientific) pp~363--367
({\it Preprint} arXiv:0708.0113 [astro-ph])

\bibitem{dm-modified} Lahanas~A~B, Mavromatos~N~E and Nanopoulos~D~V 2007 Smoothly evolving Supercritical-String Dark Energy relaxes Supersymmetric-Dark-Matter Constraints {\it Phys.\ Lett.} B {\bf 649} 83--90 ({\it Preprint} arXiv:hep-ph/0612152)

\bibitem{lahanas} Diamandis~G~A, Georgalas~B~C, Lahanas~A~B, Mavromatos~N~E and Nanopoulos~D~V 2006 Dissipative Liouville cosmology: A case study {\it Phys.\ Lett.} B {\bf 642} 179--186 ({\it Preprint} arXiv:hep-th/0605181) \\
Lahanas~A~B, Mavromatos~N~E and Nanopoulos~D~V 2007 Dilaton and off-shell (non-critical string) effects in Boltzmann equation for species abundances {\it PMC Phys.} A {\bf 1} 2
({\it Preprint} arXiv:hep-ph/0608153)

\bibitem{mavro} Mavromatos~N~E, Sarkar~S and Vergou~A 2011 Stringy Space-Time Foam, Finsler-like Metrics and Dark Matter Relics {\it Phys.\ Lett.} B {\bf 696}  300--304 ({\it Preprint} arXiv:1009.2880 [hep-th]) \\
Mavromatos~N~E, Mitsou~V~A, Sarkar~S and Vergou~A 2010 Stochastic Finsler D-particle Space-Time Foam Enhances Dark Matter Relics {\it Preprint} arXiv:1012.4094 [hep-ph]

\bibitem{dutta} Dutta~B, Gurrola~A, Kamon~T, Krislock~A, Lahanas~A~B, Mavromatos~N~E and Nanopoulos~D~V 2009 Supersymmetry Signals of Supercritical String Cosmology at the Large Hadron Collider {\it Phys.\ Rev.} D {\bf 79} 055002 ({\it Preprint} arXiv:0808.1372 [hep-ph])

\bibitem{thesis} Mitsou~V~A 2002 Prospects on Higgs boson discovery and physics beyond the standard  model at the LHC and development of the transition radiation tracker of the ATLAS experiment {\it PhD Thesis} CERN-THESIS-2002-005 \\
Mitsou~V~A 1999 Observability of the $h\to b\bar{b}$ channel in cascade decay of SUSY particles within the SUGRA model {\it ATLAS Note} CERN-ATL-PHYS-99-017 

\bibitem{mm2} Mavromatos~N~E and Mitsou~V~A 2008 Observational Evidence for Negative-Energy Dust in Late-Times Cosmology {\it Astropart.\ Phys.} {\bf 29} 442--452
({\it Preprint} arXiv:0707.4671 [astro-ph]) \\
Mavromatos~N~E and Mitsou~V~A 2007 Relaxation dark energy in non-critical string cosmologies and astrophysical data {\it Proc.\ IDM 2006: 6th Int.\ Workshop on the Identification of Dark Matter (Island of Rhodes)} ed M Axenides {\it et al} (Singapore: World Scientific) pp~623--634 ({\it Preprint} arXiv:astro-ph/0611788)

\bibitem{mm3} Mitsou~V~A 2010 Constraining super-critical string/brane cosmologies with astrophysical data {\it J.\ Phys.\ Conf.\ Ser.} {\bf 203} 012054 ({\it Preprint} arXiv:0909.5095 [astro-ph.CO])

\end{thebibliography}
\end{document}